\documentclass[preprint]{elsarticle}

\usepackage{gensymb}
\usepackage{multirow}

\usepackage{float}
\usepackage[latin1]{inputenc}   
\usepackage{amsmath} 
\usepackage{epstopdf}           %
\usepackage{flushend}           %
\newcommand{\un}[1]{\ensuremath{\,\mathrm{#1}}}
\newcommand{\degC}[0]{\ensuremath{\degree\mathrm{C}}}
\usepackage[a4paper, total={6in, 9.5in}]{geometry}
\usepackage{lmodern}
\usepackage[T1]{fontenc}
\usepackage{xcolor}

\usepackage{amssymb}

\usepackage{lineno}

\usepackage[figuresright]{rotating}
\usepackage{color,soul}

\usepackage{hyperref}
\hypersetup{
    colorlinks=true,
    linkcolor=blue,
    filecolor=magenta,      
    urlcolor=cyan,
    pdftitle={Overleaf Example},
    pdfpagemode=FullScreen,
    }

\usepackage{subfigure}

\begin{document}
\begin{frontmatter}

\title{Extensive 3D Mapping of Dislocation Structures in Bulk Aluminum}

\author[First]{C. Yildirim}\corref{cor1}
\ead{can.yildirim@esrf.fr}
\author[Second]{H.F. Poulsen}
\author[Third]{G. Winther}
\author[First]{C. Detlefs}
\author[Fourth,Fifth]{P.-H. Huang}
\author[Fourth,Fifth]{L. E. Dresselhaus-Marais}\corref{cor2}
\ead{leoradm@stanford.edu}
\cortext[cor1]{Corresponding Author}
\cortext[cor2]{Corresponding Author}

\address[First]{European Synchrotron Radiation Facility, 71 Avenue des Martyrs, CS40220, 38043 Grenoble Cedex 9, France.}
\address[Second]{Department of Physics, Technical University of Denmark, 2800 Kgs. Lyngby, Denmark.}
\address[Third]{Department of Mechanical Engineering, Technical University of Denmark, 2800 Kgs. Lyngby, Denmark.}
\address[Fourth]{Department of Materials Science \& Engineering, Stanford University, 476 Lomita Mall, Stanford, CA 94305, USA.}
\address[Fifth]{SLAC National Accelerator Laboratory, 2575 Sand Hill Rd., Menlo Park, CA 94025, USA.}
\address[Sixth]{Lawrence Livermore National Laboratory, 7000 East Ave., Livermore, CA 94550, USA.}
\date{\today}

\begin{abstract}
Thermomechanical processing such as annealing is one of the main methods to tailor the mechanical properties of materials, however, much is unknown about the reorganization of dislocation structures deep inside macroscopic crystals that give rise to those changes. Here, we demonstrate the self-organization of dislocation structures upon high-temperature annealing in a mm-sized single crystal of aluminum. We map a large embedded 3D volume ($100\times300\times300$ $\mu $m$^3$) of dislocation structures using dark field x-ray microscopy (DFXM), a diffraction-based imaging technique. Over the wide field of view, DFXM's high angular resolution allows us to identify subgrains, separated by dislocation boundaries, which we identify and characterize down to the single-dislocation level using computer-vision methods. We demonstrate how even after long annealing times at high temperatures, the remaining low density of dislocations still pack into well-defined, straight dislocation boundaries (DBs) that lie on specific crystallographic planes. In contrast to conventional grain growth models, our results show that the dihedral angles at the triple junctions are not the predicted 120$\degree$, suggesting additional complexities in the boundary stabilization mechanisms. Mapping the local misorientation and lattice strain around these boundaries shows that the observed strain is shear, imparting an average misorientation around the DB of $\approx 0.003-0.006 \degree{}$. 
\end{abstract}


\end{frontmatter}


\section{Introduction}
Since dislocations were first postulated as the lattice defects responsible for the plastic deformation and workability of metals \cite{Orowan1934,Taylor1934,Polaniy1934}, their behavior has been an active field of research \cite{ Anderson2017}. Extensive work has resolved that dislocations spatially organize (i.e. pattern) during plastic deformation into hierarchical networks. Dislocation networks pack into 3D structures that comprise grain and domain boundaries that separate nearly dislocation-free cells in the crystal; these networks define the microstructure and affect a metal's mechanical properties. Transmission electron microscopy (TEM) studies have formulated trends for how dislocations pack to distort the structure of materials, and have developed scaling laws to relate the distribution of the distances between boundaries \cite{Godfrey2000} to the crystallographic misorientation they accommodate \cite{Hughes1998}. 
Empirical relations have formulated a more fundamental view, relating the misorientation across their boundaries, the morphology of each dislocation that comprises the boundaries, and their Burgers vectors and glide planes to determine the selection rules that govern how dislocations can pattern during crystal deformation \cite{Winther2007, Le2012}. 
In a limited number of cases, the characters of the dislocation networks in the boundaries \cite{McCabe2004,Wei2011,Winther2015} have been determined. 
Despite this progress in characterization, the mechanisms governing the dynamics in this patterning are still poorly understood. 

Until recently, the prevalent technique used to study dislocation patterning has been TEM, which resolves dislocations by imaging through thin foils ($\sim200$nm). Studies of foils are not necessarily representative of the bulk because the dislocations' attraction to surfaces can alter their interactions \cite{hata2020electron, feng2020tem}. 
Deformation-induced planar boundaries typically form along nearly parallel planes and extend over several tens of micrometers. 
More importantly, the spacing between these boundaries is often much larger than the foil thickness, meaning that patterning is only clearly observable when the foil is cut along a plane nearly orthogonal to the normal vectors to the boundaries. 
As such, the low dislocation density structures present in annealed single crystals are virtually impossible to capture and characterize in thin TEM foils, even though they are essential for understanding how and why dislocations pack into their lowest energy structures. 

By contrast, Dark-Field X-ray Microscopy (DFXM) is a new, synchrotron-based imaging technique that is conceptually similar to dark-field TEM but with a deeper penetration and a higher angular resolution that is afforded by the X-ray objective, which is placed along the diffracted beam \cite{Simons2015,Poulsen2020}. 
With a field of view of several hundreds of micrometers, 3D characterization of large volumes can be achieved by scanning and stacking adjacent layers, in a "section-DFXM" scan. 
The application of X-rays to study dislocations is not new. 
Conventional topography \cite{Tanner1976} and synchrotron based methods such as topo-tomography \cite{Ludwig2001} and laminography \cite{Hanschke2012} can also provide a large field of view. 
With a spatial resolution of 2-10 $\mu$m, these methods are well suited for semiconductor single crystals. 
In comparison to these methods, DFXM reaches a substantially higher spatial resolution on the microstructure,  as the objective both magnifies the image and separates the angular- and direct-space information. This allows for identification of 3D dislocation boundaries with high resolution in both strain and grain orientation.

Initial DFXM studies have focused on the evolution of the microstructure in metals during recovery and recrystallization \cite{Simons2015, Ahl2017, Mavrikakis2019, yildirim20224d}, as well as domain evolution in ferroelectrics  \cite{Simons2018}, and martensitic phase transformations \cite{Bucsek2019}. 
In 2019, DFXM-based methods were presented to map individual dislocations -- and demonstrated on examples including threading dislocations in SrTiO$_3$ and misfit dislocations in a BiFeO$_3$ film \cite{Jakobsen2019, Simons2019}. 
As is often described in TEM \cite{cockayne1969}, Jakobsen et al.  demonstrated with DFXM how \emph{weak-beam contrast}  describes images collected with the sample oriented to diffract only at the most highly-deformed regions of the material (i.e. images recorded at the tails of the rocking curve), namely, those surrounding a defect core.
Their work demonstrated the utility of weak-beam contrast in capturing dislocation lines with high specificity and spatial clarity, giving a clear 2D view of the projected dislocation structures.

In this paper, we use the novel technique, DFXM, to study the deeply-embedded 3D dislocation boundaries that form under the lowest-energy configurations. 
We characterize the assembly of dislocation boundaries by measuring the structures formed from high-temperature annealing in initially undeformed crystals. 
Under these conditions, the dislocations are free to self-organize into their \emph{preferred configurations}, without being restricted by the kinetic or spatial features present in the higher dislocation-density systems studied previously \cite{humphreys_2004}. 
With DFXM, we map dislocations over a large 3D volume ($100\times300\times300$ $\mu$m$^3$) of a single-crystal of aluminum, annealed at a temperature close to the melting point ($0.9T_m$). 
While this system is related to the previous studies on deformed and annealed polycrystals \cite{Godfrey2000}, the dislocations in the single crystals presented in this work are distinct: they are not hindered by stacked dislocation or grain boundaries, affording long mean free paths that facilitate motion. 
Our different but related view of dislocation organization provides key insights into the energy landscape of the reorganization process. 
Using DFXM, we resolve the individual dislocations in each boundary and demonstrate how the boundaries pack over long length-scales; based on the low-energy configurations, we show how localized distortions instill preferential patterning and organization to stabilize locally irregular defects.

\section{Experimental Methods}
\subsection{Samples}
The sample used in this work was a single crystal of industrially-pure aluminum (99.99\%), purchased from the Surface Preparation Laboratory, with dimensions $0.7\times 0.7 \times 10\un{mm^3}$, oriented with the long $[1\bar{1}0]$ axis perpendicular to the scattering plane. Prior to the experiment, the sample was annealed for 10 hours at 590$\degC$, then slowly cooled in the furnace.

\subsection{DFXM}
\begin{figure}[!h]
    \centering
    \includegraphics[width = 0.98\textwidth]{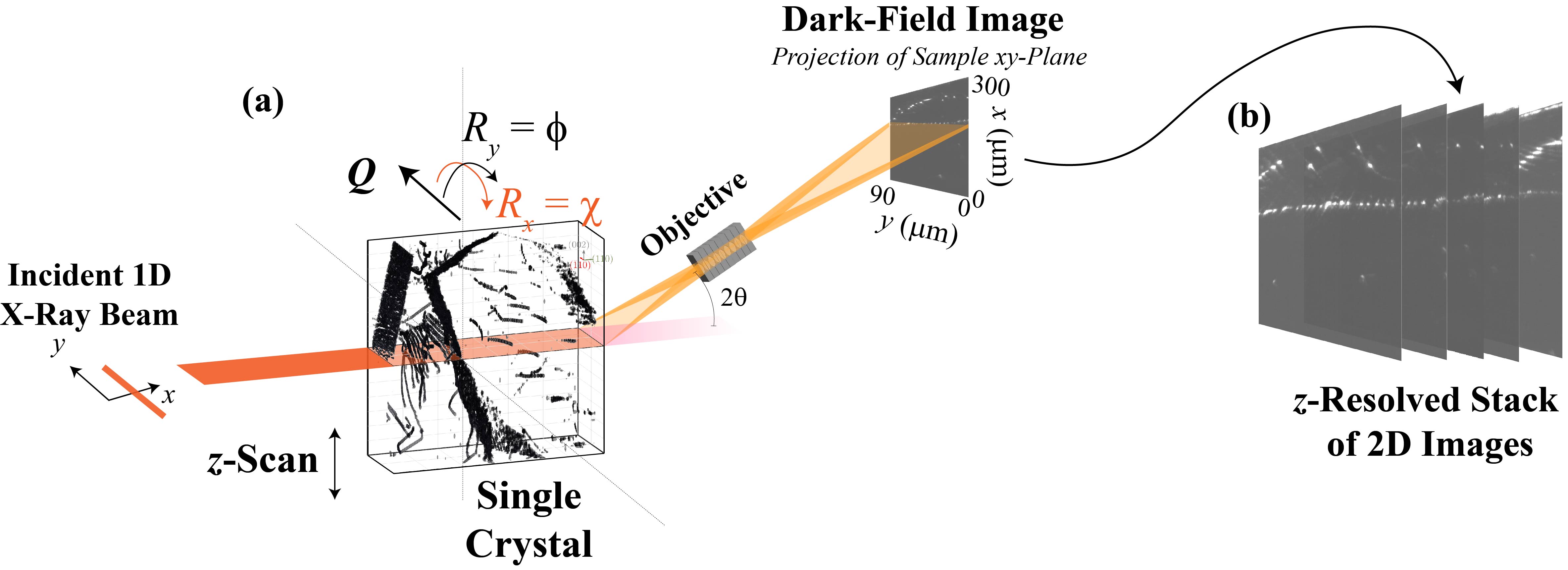}
    \caption{Schematic showing the DFXM experiment and how it captures each image in our 3D reconstruction, as plotted in the laboratory coordinate system. The \emph{observation plane} for each image shown by the orange plane that slices through the cube showing the dislocations we observed in our single crystal. The direction of the scattering vector $\mathbf{Q}$ can be varied by the two tilts, $\phi$ and $\chi$. The length of the scattering vector can be varied by a combined  $2\theta-\phi$ scan.}
    \label{fig:Schematic}
\end{figure}

Our DFXM experiments were conducted at Beamline ID06-HXM at the European Synchrotron Radiation Facility (ESRF) \cite{Kutsal2019}. We used 17\un{keV} photons, selected by a Si (111) Bragg-Bragg double crystal monochromator, with a bandwidth of $\Delta E/ E= 10^{-4}$. The beam was focused in the vertical direction using a Compound Refractive Lens (CRL) comprised of 58 1D Be lenslets with an R=100\un{\mu m} radius of curvature, yielding an effective focal length of 72\un{cm}. The beam profile on the sample was approximately $200 \times 0.6 \un{\mu m^2}$ (FWHM) in the horizontal and vertical directions, respectively. The horizontal \textit{line beam} illuminated a single plane that sliced through the depth of the crystal, defining the microscope's \textit{observation plane}, as shown in Fig.~\ref{fig:Schematic}. A  near-field alignment camera was placed 40\un{mm} behind the sample, and used to orient the crystal into the Bragg condition.
Following alignment, the near-field camera was removed and the image was magnified by an X-ray objective lens comprised of 88 Be parabolic lenslets (2D focusing optics), each with a R=50\un{\mu m} radii of curvature. The entry plane of the imaging CRL was positioned 281\un{mm} from the sample along the diffracted beam, and aligned to the beam using a far-field detector. The objective projected a magnified image of the diffracting sample onto the far-field detector, with an X-ray magnification of $M_{\text{x}}= 17.9$$\times$. 

Our far-field imaging detector used an indirect X-ray detection scheme, using an zoom to impart additional magnification. This detector was comprised of a scintillator crystal, a visible microscope and a $2160 \times 2560$ pixel PCO.edge sCMOS camera. It was positioned 5010\un{mm} from the sample. The visible optics inside the far-field detector could switch between $10\times$ and $2\times$ objectives to achieve an effective pixel size of 0.75\un{\mu m} or 3.75\un{\mu m}, respectively. This paper focuses on analysis from the highest-magnification 10$\times$ magnification images (total magnification of $M_t=179\times$). 

This study focuses on dislocation structures observed by diffraction contrast microscopy from the $(002)$ Bragg reflection, with a corresponding diffraction angle of $2\theta = 20.77$\degree{}. To obtain 3D information, DFXM images were collected for 2D layers, scanning the sample in the vertical direction, $z$, to resolve variation along the height of the crystal (as shown in Fig.~\ref{fig:Schematic}a). 

Three types of scans were performed in this work: rocking scans, mosaicity scans, and axial strain scans. The rocking scans acquired images while scanning the tilt angle $\phi$, see Fig.~\ref{fig:Schematic}, over a range of  $\Delta \phi = 0.12\degree{}$ 
in 30 steps (i.e. $\delta \phi = 0.004\degree{}$ per step). These 1D ``scans of the rocking-curve'' map components of the displacement gradient tensor field (i.e. strain and misorientation) to indicate the local variation in structure that is relevant to visualize dislocations \cite{Poulsen2021}. We collected this type of data for a total of 301 layers, with  1-$\mu$m steps between the layers. The resulting information in four dimensions $(x,y,z,\phi)$ was imported into Matlab for subsequent processing and feature identification.

Additional supporting scans were collected to more thoroughly sample the mosaicity for selected layers, by measuring distortions along the two orthogonal tilts $\chi$ and $\phi$, cf. Fig.~\ref{fig:Schematic}. The $\chi$-range and step size was $\Delta \chi = 0.24 \degree{}$  and $\delta \chi = 0.024\degree{}$, respectively, while the $\phi$-range and step size was the same as for the rocking scans. With this data, each voxel can be associated with a subset of a $(002)$ pole figure, allowing us to generate Center of Mass (COM) maps to describe the average direction of the $(002)$ orientation for each voxel in the layer \cite{darfix}. We note that the angular resolution in the COM maps is substantially better than the step size. Finally, axial strain scans were collected by keeping all orientations fixed, while scanning the $2\theta$ axis to resolve the axial strain component $\varepsilon_{33}$, then reconstructed into the same COM map to quantify the residual strain in each voxel.

\subsection{Data Analysis Methods}
All dislocations were identified from the  $z_{\ell}$-resolved stack ($\{x_{\ell},y_{\ell},z_{\ell}\}$ is the laboratory coordinate system) of $\phi$-resolved rocking scans using thresholding and a fiber-texturing methods for reconstruction. We first reduced the 4D ($x,y,z,\phi$) dataset into an entirely spatial ($x,y,z$) 3D dataset by manually identifying the $\phi$ position characteristic of the weak-beam condition for each 2D layer (i.e. with the brightest and most sharply-defined spots for dislocations). The ideal $\phi$ orientation changed only slightly over the course of overnight scans -- likely because of drift in the microscope or continuous rotation of the grain's orientation (i.e. bending). For the 3D image stack, intensities below a defined noise-floor were set to the threshold value of $I_{min}=110$, and the edge 10 pixels were also set to $I=110$ to avoid deleterious edge effects from the detector. The resulting stack of raw 2D images (i.e. Fig.~\ref{fig:linefig}b) captures the features that appear in the weak-beam condition and set a uniform distribution of pixel intensities for each image when compared to the other layers, enabling for subsequent 3D processing. Each ($x,y$) image was then 2$\times$2 binned before being saved into a 3D image cube (binning was necessary to reduce memory usage). The image cube was then input into the {\fontfamily{qcr}\selectfont
fibermetric()} function in Matlab to connect the diffuse intensity corresponding to linear dislocation lines. This function filters the image by identifying the voxels that are characteristic of bright linear features in the 3D image arrays, based on a gradient method (set to a Structure Sensitivity of 5). The resulting map was plotted using the {\fontfamily{qcr}\selectfont PATCH\_3Darray()} surface visualization method \cite{SurfTool}. 

After defining the regions characteristic of each boundary, we indexed the dislocation boundary planes by defining all relevant points characteristic of each plane as a {\fontfamily{qcr}\selectfont pointCloud} object. For each boundary, we identified the local normal vectors that are characteristic of each voxel based on the 10 nearest neighbors, using the {\fontfamily{qcr}\selectfont pcNormals()} function \cite{Hoppe1992}. The estimates from this {\fontfamily{qcr}\selectfont pointCloud} were then input to the  {\fontfamily{qcr}\selectfont pcfitplane()} method in Matlab, which uses an M-estimator SAmple Consensus (MSAC) to find the plane (related to the RANdom SAmple Consensus, RANSAC approach), which excludes outlier points and outputs a mean uncertainty value \cite{torr2000mlesac}. The resulting plane fits gave mathematical uncertainty quantification, based on the mean error in $\mu$m that points lie beyond the boundary place, $\delta$z$_{err}$. We input the Cartesian vectors estimated by MSAC to a transformation matrix to converts the vectors described in detector-frame into our coordinate system that takes into account the crystallographic basis vectors (i.e. $\vec{a}_1$, $\vec{a}_2$, $\vec{a}_3$). Because $\theta = 10.38^{\degree}$, our transform matrix multiplied the crystal's native orientation vectors by the $\theta$ rotation about the $y$ axis in the detector plane (i.e. vertical diffraction) to account for the angle between the observation plane that defines each image and the $(002)$ diffraction plane that sets the contrast mechanism. We then transformed the $[x,y,z]$ vectors into their associated $[uvw]$ vectors in the lattice system, then inverted them to solve for the $[hkl]$ vectors necessary for our interpretation. After rounding the appropriate values, we re-plotted the normal vectors and associated planes to verify the accuracy, as described fully in the Supplemental Information. We note that since MSAC is a random-sampling function, the method does not give a deterministic output; as such, we monitored the output many times until converging on a solution that fit the points accurately even after rounding errors. 

Dimensional reduction was then performed a second time to identify the line vectors that were characteristic of each dislocation. Using the newly defined boundary plane normal, two orthonormal vectors were identified that lie within the DB, and an affine transform was defined to convert the boundary points from the $\{x_{\ell},y_{\ell},z_{\ell}\}$ laboratory coordinate system into the newly defined  $\{x_{\text{bp}},y_{\text{bp}},z_{\text{bp}}\}$ boundary plane coordinate system. In this system, $\hat{z}_{\text{bp}}$ was defined as the unit vector describing plane normal, allowing all points to be reduced from their 3D representations in ${\rm I\!R}_{\ell}$ to their 2D representations in ${\rm I\!R}_{\text{bp}}$. The 2D points were then rotated via a new affine transform to express the boundary as a 1D system showing the position of each dislocation, i.e. the dislocation system ${\rm I\!R}_{\text{d}}$. To identify the appropriate angle to convert the 2D boundary plane system into the 1D dislocation system, a sequence of 180 candidate transformations were assembled by scanning the rotation matrix through all unique 180$^{\circ}$ angles, then compiling the rotated points into a histogram of counts vs $x_{\text{d}_i}$. The Fourier transform from each trace was plotted as a function of angle, and the ``aligned'' angle was identified as the point for which the number of spatial frequencies required to describe the 1D function was most sharply defined (as opposed to the scattered and diffuse frequency components required to describe the misaligned ones). All assignments were confirmed graphically, as demonstrated in the Supplemental Information.

We include the Matlab scripts and functions used in this work in the Github folder available at \url{https://github.com/leoradm/Dislocation3DAnalysis.git}.


\section{Results}

\begin{figure}[h!]
    \centering
    \includegraphics[width=15cm]{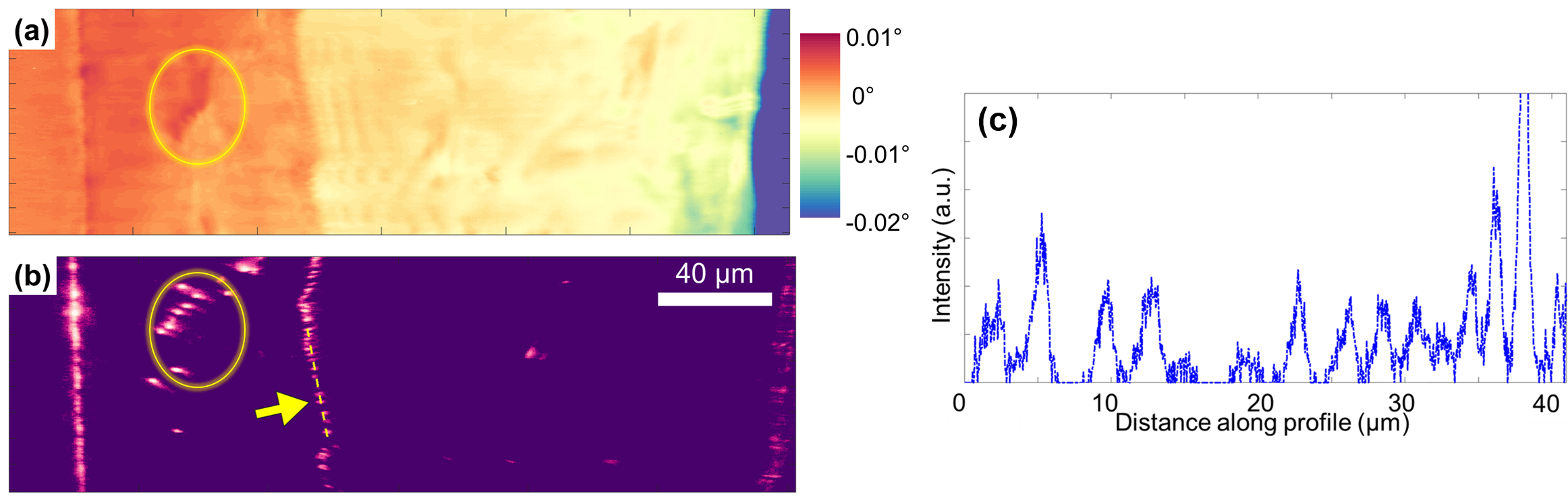}
    \caption{(a) Computed center of mass map of sample rotation in $\phi$  of a selected  slice through the thickness, with colors defined by the colormap to the right.  A full movie showing the $\phi-\chi$ mosaicity map over 40 $\mu$m in $z$ direction with 2 $\mu$m steps is presented in Supp. Mat. (b) Weak-beam contrast of the same slice showing dislocation boundaries as the pink dotted lines, like the one circled in yellow, for clarity. (c) The intensity profile of the dashed-line trace in (b), showing peaks that describe the positions of each dislocation in the boundary (B1 in Fig. \ref{fig:Boundary1}).}
    \label{fig:linefig}
\end{figure}

We begin by showing results for a mosaicity scan from a single layer.  Fig.~\ref{fig:linefig}a presents the $\phi$ rotation COM map (i.e. rocking curve COM map \cite{darfix}).  
In this COM map, we see three primary subgrains (orange, yellow, and blue) that are separated by boundaries that appear as nearly vertical lines that discontinuously change the local orientation. By comparing the COM image in Fig.~\ref{fig:linefig}a to a single image from the same layer that satisfies the weak-beam condition from the ($\phi,\chi$) scan in Fig.~\ref{fig:linefig}b, we see that the orientational shifts across the boundaries correspond to apparently dotted lines in the raw image. The dotted lines shown with the yellow arrow indicating an array of discrete dislocations whose line vectors are steeply inclined with respect to the observation plane \cite{Poulsen2021}. The COM map shows that all three subgrains have rather homogeneous angular spreads, except in locations that have internal dislocations within the cell, as indicated in Fig.~\ref{fig:linefig}a-b by the yellow circle. These are localized areas of strong intensity, characteristic of the strain field of isolated dislocations. We focus on the row of dislocations with an overlaid yellow dashed line that separates the yellow and orange subgrains. For this boundary, we demonstrate the spacing between the dislocations by plotting the intensity-trace depicted by the yellow line in Fig.~\ref{fig:linefig}b as a plot in Fig.~\ref{fig:linefig}c. The average distance between the dislocations along the boundary is $D=$ 4.1 \un{\mu m}. This misorientation across the boundary as determined from the COM map is $\Delta \phi$ = 0.004\degree, while the Burgers vector has a magnitude of $b$ =2.86 \AA.

We note that classical dislocation theory predicts a misorientation of $\Delta \phi=b/D$ for a dislocation boundary of Burgers vector, $b$, and spacing, $D$ \cite{Hull2011}. Our measurement of the crystal misorientation across the boundary and the corresponding dislocation spacing we measure from our weak-beam image fit this model precisely.

From the 1D illumination in Fig.~\ref{fig:linefig}b, the small dots for each dislocation indicate that each dislocation line slices through the 2D observation plane defined by the 1D X-ray line-beam illumination. As demonstrated in Supplementary Material, the 3D position of dislocation lines cannot be traced simply by making the incident beam larger. Instead, we compile a spatial 3D map of dislocations by stacking the results for the individual layers of the kind shown in  Fig.~\ref{fig:linefig}b.

\begin{figure} [h!]
    \centering
    \includegraphics[width=10cm]{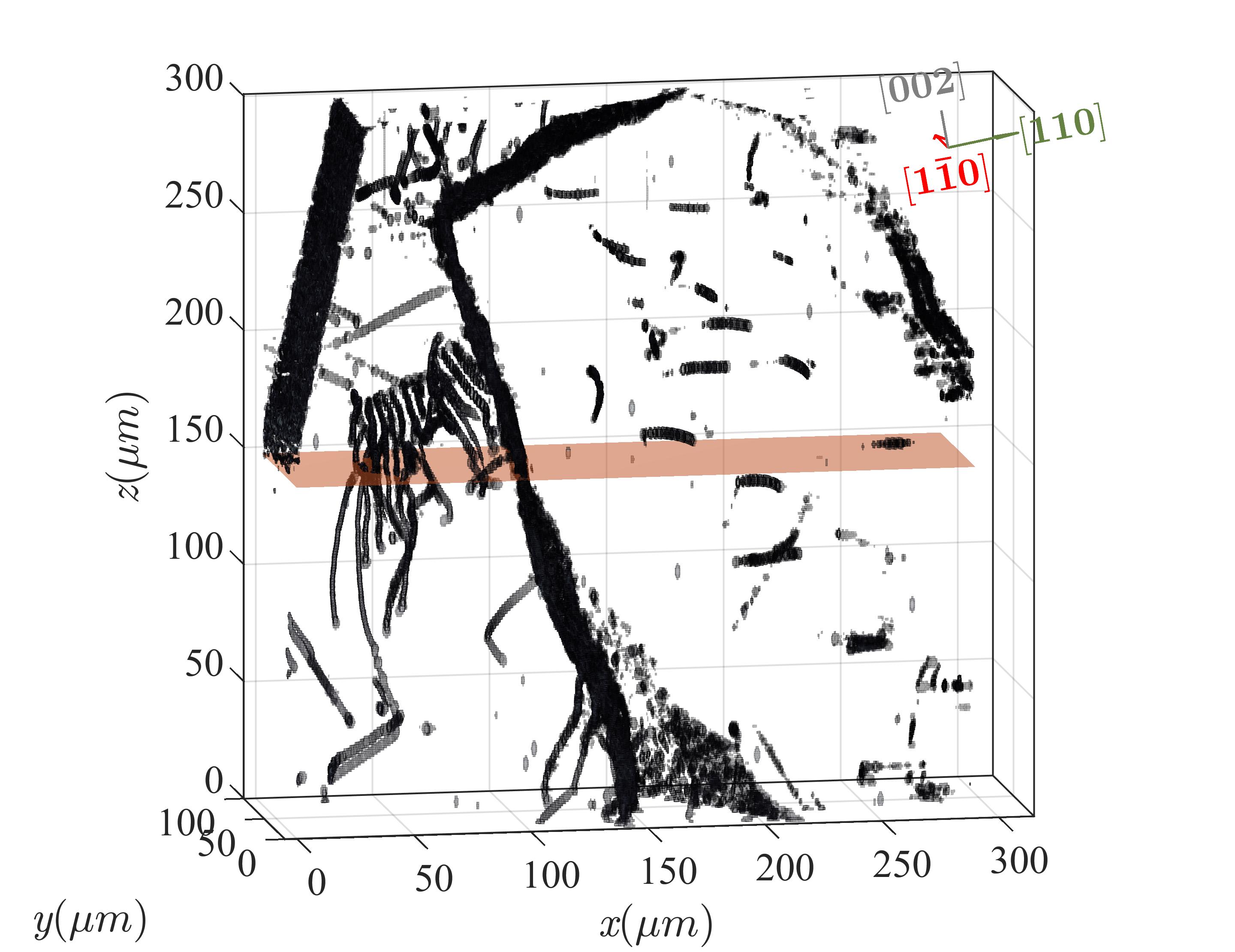}
    \caption{3D map of dislocation structures over a 100$\times$300$\times$300-$\mu$m$^3$ volume. All dislocation features are shown by the grayscale surfaces, with the crystallographic coordinate axes inserted in the top right of the plot. Each voxel in the map is $125 \times 45 \times 1000$ nm$^{3}$. The orange plane represents the 2D slice shown in Fig~\ref{fig:linefig}. Note that the $\chi$ position of the scan shown in Fig~\ref{fig:linefig} is $0.02\degree$ different than Figure \ref{fig:3Ddislocations}.} 
    \label{fig:3Ddislocations}
\end{figure}

The resulting 3D dislocation structures resolved with our section-DFXM approach are shown in Fig.~\ref{fig:3Ddislocations} for the full volume probed at the highest magnification. Fig.~\ref{fig:3Ddislocations} shows that dislocations in the probed 3D volume self organize into preferential structures. The map comprises  clearly defined lines that are analogous to those seen via dark-field TEM at smaller scales: they represent the dislocation lines, as measured by the locally high strain and orientation components that become asymptotic immediately surrounding the core structure \cite{Poulsen2021}. The dislocations identified in this volume clearly pack in hierarchical structures: a large collection of dislocations is present, and furthermore, in some cases, the dislocations pack into long-range boundary structures (e.g. those on the right) that separate different subgrains of the crystal. To understand the mechanics of a crystal at the mesoscale, we explore different types of dislocation packing arrangements in the crystal, interpreting key details of the boundaries at the scale of the boundary planes and the component dislocations. 

\begin{figure}[h!]
    \centering
    \includegraphics[width=15cm]{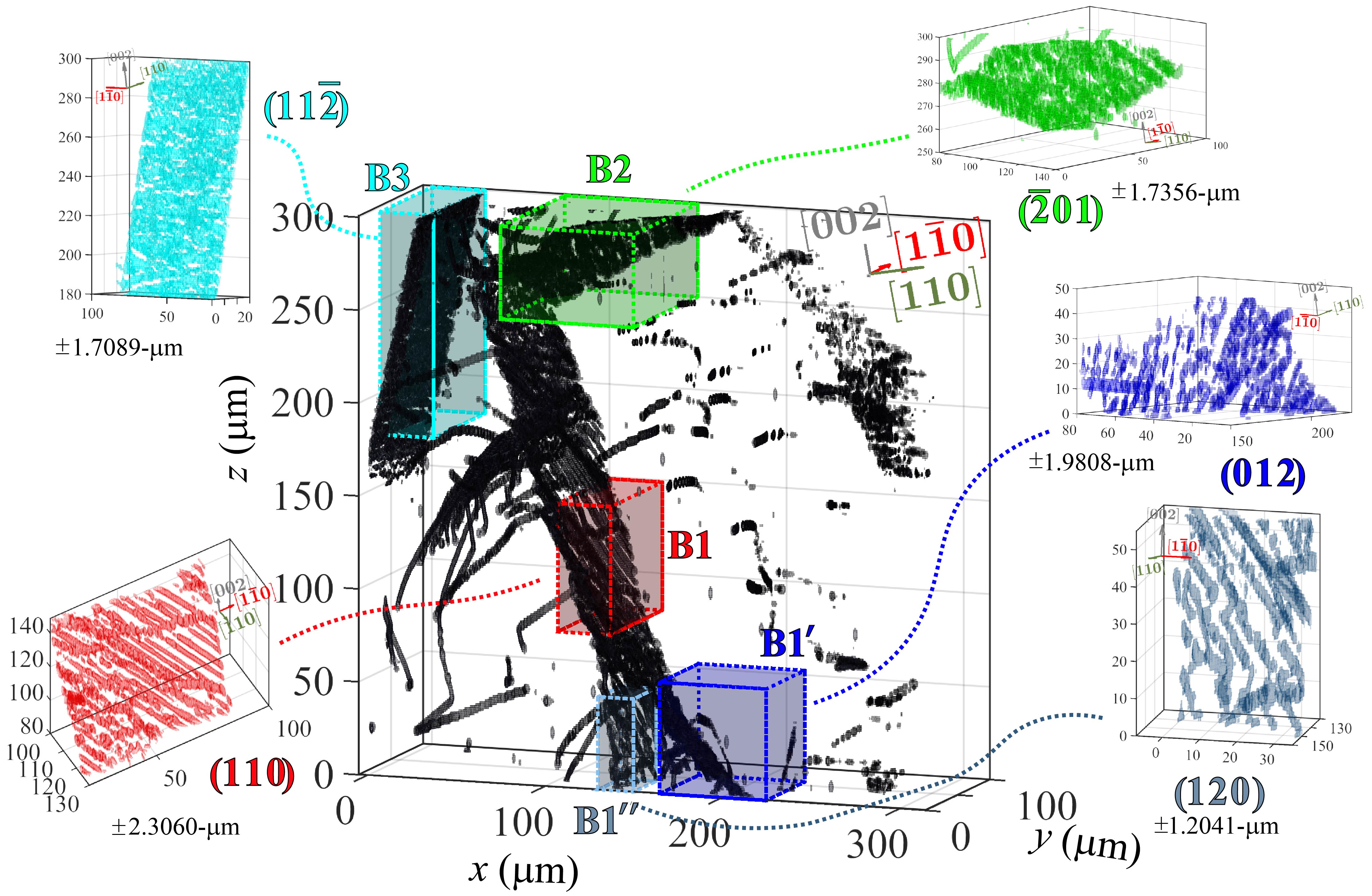}
    \caption{Labels showing the five primary dislocation structures shown in Fig.~\ref{fig:3Ddislocations}, each labeled and outlined in a different color. Shown as inlets are zoom-ins on each boundary to demonstrate its orientation, and demonstrate with the ``zoomed-in'' view how the single dislocations in that boundary are packed. Mean error values from the MSAC fitting is listed for each plane in the respective plots. }
    \label{fig:Boundary1}
\end{figure}

We show an annotated version of the 3D dislocations from Fig.~\ref{fig:3Ddislocations} in Fig.~\ref{fig:Boundary1} to present in detail the structure within five crystalline regions (identified by five colored boxes) that are characteristic sections of each well-defined dislocation boundary (DB) we describe in this work. A clear picture now emerges on the self-organization of the dislocation structures in the probed volume. We observe that dislocations pack along well-defined planes within 3D, even after long annealing times at temperatures close to melting. From a first glance, the dihedral angles of the triple junctions are far from
120\degree, in contrast to what conventional growth models predict. Below, we zoom in on individual boundaries shown in Fig.~\ref{fig:Boundary1} and analyze in detail to extract more information on the self-organization process. 
Inlays in Fig.~\ref{fig:Boundary1} obtained from zooming in on each separate low-angle boundary plane demonstrate that we can resolve full structures. In particular, we can resolve the defect plane that separates sub-domains of the crystal, and by projecting each boundary along different vectors we can identify the relevant in-plane and out-of-plane directions crystallographically. As described above, this allowed us to solve for the zone axes for the 5 primary boundaries in this structure, as labeled in Fig.~\ref{fig:Boundary1}. To further refine our assignments, we then isolated each DB and viewed each one along the possible zone-axis vectors, verifying that the appropriate vector corresponds to the one with the largest spatial extent within our view, and two normal vectors that constrain the plane to linear features.
Table ~\ref{table:boundaries} show the results for all of the boundaries in Fig.~\ref{fig:Boundary1}.  

\begin{center}
\begin{table}
\caption{Summary of the fitted boundary plane normal vectors ($\vec{n}$), the calculated Burgers vectors ($\vec{b}$), and the measured dislocation line vectors ($\vec{t}$) for each boundary plane, denoted with the same labels from Fig. \ref{fig:Boundary1}. The  mean error ($\delta$z$_{err}$) from the plane fitting is also given. $\gamma$ is the average misorientation angle ($\Delta \phi$) measured across the boundary (extracted from the measured $\phi$ scans).}
\vspace{2mm}
\centering
    \begin{tabular}{ c c c c c }
        \bf{Boundary} & \bf{$\vec{n}$} (hkl)  & \bf{$\vec{t}$} (hkl) &   \bf{$\delta$z$_{err}$ ($\mu$m)} & \bf{$\gamma$ ($\times 10^{-3}$)} \\ \hline
        B1 & [$110$] &  [$2\bar{2}1$]  & 2.3006 & 0.0418$^{\circ}$ \\  
        B1$'$ & [$012$] & [$2\bar{2}1$] &  1.9808 & 0.2370$^{\circ}$ \\
        B1$''$ & [$120$]  &[$3\bar{2}1$] &  1.2041 & 0.2332$^{\circ}$ \\
        B2 & [$\bar{2}01$] &  [$1\bar{2}2$] & 1.7356 & -0.0284$^{\circ}$ \\
        B3 & [$11\bar{2}$] & [$\bar{1}10$]  & 1.7089 & 0.2493$^{\circ}$ \\  
    \end{tabular}
\centering
\label{table:boundaries}
\end{table}
\end{center}

\subsection{Characterization of Triple Junction: B1-B1$'$-B1$''$}
Boundary B1 (red) is the primary boundary that slices down the middle of the characterized volume; the $hkl = [110]$ normal vector defines the B1 plane. The inlay in Fig.~\ref{fig:Boundary1} shows that B1 is comprised of clearly defined, straight linear features that indicate the direction of the dislocation lines in the boundary.

\begin{figure}[!h]
    \centering
    \includegraphics[width=15cm]{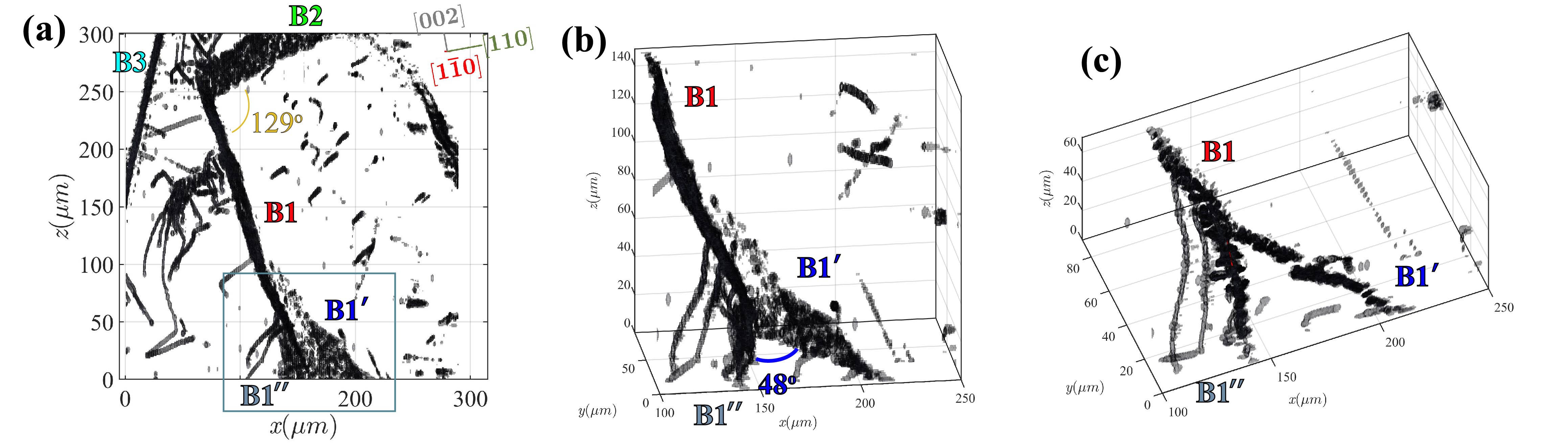}
    \caption{(a) Projection of the 3D dislocation structures shown in Fig.~\ref{fig:Boundary1} along the crystallographic $[1\bar{1}0]$ vector, showing the angles between each boundary plane. With this perspective, the space between B3 and the intersection of B1 and B2 is clearer, as well as the curvature of the boundaries when B1 and B1$'$ split, near the bottom. (b) A zoomed-in and rotated perspective of the junction between B1 and B1$'$, showing the triple-junction formed by the two planes adjoining. Angle between planes is labeled. (c) Projection of the boundary off-crystallographic axis, showing the curvature and discontinuity of B1$'$  and the component dislocations.}
    \label{fig:2DProjection}
\end{figure}

A closer look at the bottom of Fig.~\ref{fig:2DProjection}(a) shows that B1 forms a triple-junction with boundaries B1$'$ and B1$''$. The boundary dislocations in B1 and B1$'$ bend around the junction point, making both become curved planes over a region. B1$'$ discretely changes at a ``kink point'' that makes the boundary flatten into a planar boundary that is normal to the $[012]$ vector with straight dislocations that point along the $\vec{t}=[02\bar{1}]$ vector. The dislocations in B1$''$ gradually bend onto the new $(120)$ plane (with dislocation line vectors that primarily point along $[02\bar{1}]$, and have some possible kinks). Based on the angle between the vectors normal to the planes, the angle at the triple-junction is $37\degree$ immediately surrounding the B1$'$ and B1$''$ junction, but the bend of the B1$''$ plane ultimately shifts the long-range boundary angle to $48\degree$ as measured further away from the junction. 

The curvature of the boundaries near the triple junction suggests that the stabilizing force usually predicted from triple junctions may not be valid in this system. Instead, we interpret that the curvature surrounding this junction may arise from a localized impurity that pinned a single dislocation during the annealing process, then was stabilized by the three surrounding boundaries, B1, B1$'$ and B1$''$. The curvature in these boundaries near the junction suggests the innermost dislocation has the highest energy, thereby distorting the topology of the boundary planes at that site. 

\subsection{Boundary Identification}
Beyond B1 and B1$'$, Fig.~\ref{fig:2DProjection} shows two other boundaries that are clearly defined in the high-$z$ regions of the dislocation structures. Another boundary, B2 (green), intersects B1 in the upper $\sim$50-$\mu$m $z$ region of the volume, intersecting at an angle of $129\degree$ (Fig.~\ref{fig:2DProjection}a). B2's plane is defined by its $[\bar{2}01]$ normal vector, with dislocations that pack significantly closer together, as observed in the inlay, packed along the $[1\bar{2}2]$ vector. Near the region where the B1 and B2 planes intersect, a third boundary -- B3 (cyan) -- nears the edge of the dislocation structures, with a normal vector along $[11\bar{2}]$. The spacing between the $\vec{t}=[\bar{1}10]$ dislocations in B3 is the smallest of all five boundary structures, imposing the highest misorientation angle. As corroborated by the COM map shown for the top layer in Fig.~\ref{fig:linefig}a, the $\sim0.015\degree$ misorientation across the boundary, indicating a $\sim780$-nm spacing that makes the dislocations difficult to differentiate based on the thresholds used in our segmentation methods. We note that with higher precision afforded by Bayesian inference, future implementations of this method could improve the resolution significantly \cite{Brennan}. From Fig.~\ref{fig:2DProjection}, the projected image along the $[1\bar{1}0]$ axis shows that B3 never intersects boundaries B1 and B2. Note that some of the dislocations shown in Fig.~\ref{fig:linefig}b do not appear in these volumes as they only satisfy the Bragg condition, thus become visible, at certain $\chi$ tilts. These 3D maps generated from the 4D scans ($x,y,z,\phi$) are measured at a fixed $\chi$ value. 

Going beyond the well-formed boundaries, we also note that this weak-beam 3D DFXM scan allows us to map the isolated (lone) dislocations quite effectively as well. Fig.~\ref{fig:3Ddislocations} and Fig.~\ref{fig:2DProjection} show an interesting and complex dislocation structure between B1 and B3. This structure includes curved dislocations that appear to form a complex boundary shape with significant curvature. It is possible that this dislocation structure connects to B1 and B2, forming another triple junction. The irregular character of this dislocation structure indicates DFXM's ability to characterize structures with complexity beyond a classical boundary. For example, one of the dislocations in this structure stretches down $\approx$100-$\mu$m along $[00\bar{2}]$ direction before it truncates in a partial loop centered around $(x,y,z)=(40,40,40)$. While a precise analysis of this unusual dislocation tangle is beyond the scope of this work, we note that Section-DFXM provides a new approach to characterize these complex structures using the 3D image segmentation techniques to resolve a deeper view of complex topologies.

\subsection{Analysis of Individual Boundaries}

If a boundary is not associated with long-range stresses, the dislocation arrangement in the boundary should fulfill the Frank equation \cite{Frank1950}:

\begin{equation}
    \Sigma \rho _i\vec{b}_i\{(\vec{n} \times \vec{t}_i)\cdot\vec{r} \} = 2 sin(\gamma /2) (\vec{r} \times \vec{a}),
\end{equation}
where $\rho _i$ and $\vec{t}_i$ are are the density and line direction of the dislocation with Burgers vector $\vec{b}_i$. The boundary plane normal and misorientation axis are $\vec{n}$ and $\vec{a}$, respectively. $\vec{r}$ is an arbitrary vector that lies in the boundary plane.

All of the boundaries in the observation volume (Fig.~\ref{fig:3Ddislocations}) have straight parallel dislocation lines as the dominant feature. Some indications of crossing dislocation lines may be seen but their densities are low. A boundary consisting of dislocations of only one Burgers vector must lie on the plane with the Burgers vector as the normal \cite{Frank1950}. In fcc this implies boundary planes of the {110} family. As seen in Table 1 this is the case for boundary B1. The classical boundary of this type is a tilt boundary consisting of edge dislocations with dislocation line along <11$\bar{2}$>, which enter the boundary by glide. By contrast, the dislocation line for B1 is [2$\bar{2}$1], which does not lie in any slip plane. It can be inferred that it is the high temperature which enabled this motion out of the slip plane by climb \cite{Hull2011, zhang2010high}. 

For the rest of the boundaries, the parallel dislocations must have different Burgers vectors to fulfill the Frank equation. The Burgers vectors of each dislocation cannot be identified at present, though Table 1 shows the \textit{average} Burgers vectors for all boundaries. For B3, the boundary plane and dislocation line directions are consistent with the Frank equation as a tilt boundary, with equal densities of dislocations with Burgers vectors of [10$\bar{1}$] and [01$\bar{1}$].

For the remaining boundaries, B1$'$, B1$''$ and B2, the boundary planes and dislocation line directions are symmetrically equivalent with planes of \{012\} and line directions of $<$221$>$ to $<$321$>$. With the ($\bar{2}$01) plane of boundary B2 as an example, the Frank equation was employed to establish that a boundary on the ($\bar{5}$02) plane consisting of two sets of parallel dislocation lines along [1$\bar{4}$5] with Burgers vectors [$\bar{1}$01] and [01$\bar{1}$] fulfills the equation if the density of the first Burgers vector is larger than the density of the other by a factor of about 2.3. The angle between the experimentally observed boundary plane and the one obtained using the Frank equation is $5 \degree$ and the theoretical dislocation line lies in between those determined experimentally for the three symmetric boundaries. These deviations may be due to the presence of a small density of additional dislocations. Analogous analyses for B1$'$ and B1$''$ can be made. 

\section{Discussion}

\subsection{Dislocation boundaries}
 Upon closer examination, we can see that our 3D map with DFXM resolves the dislocation lines that comprise the boundary planes. In this case, our dislocations boundary planes are comprised of the boundary dislocations. The boundary dislocations are relatively straight across the entire spatial extent of the boundary we resolve. This detailed view of dislocation packing suggests that the boundaries manifest the lowest energy structures that may form under the high-T processing conditions. 

\begin{figure}[!h]
    \centering
    \includegraphics[width=15cm]{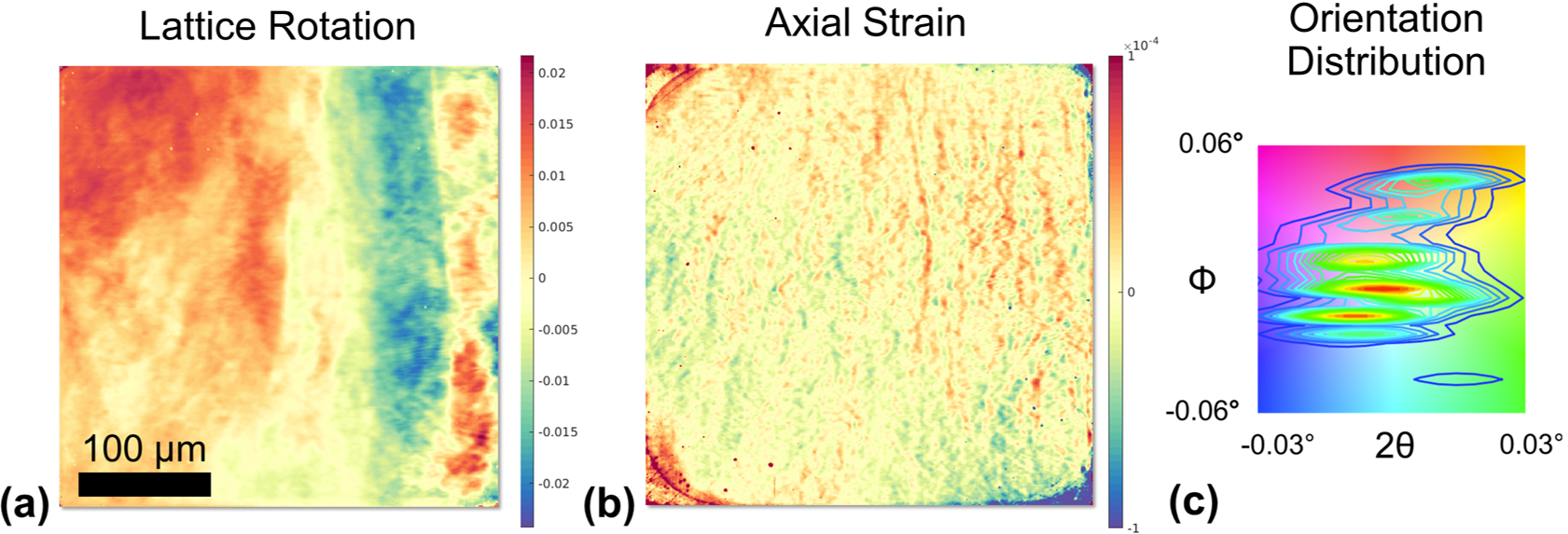}
    \caption{Computed COM maps of $\phi$ rotation (i.e. rocking curve COM map) (a) and relative axial strain (b) of (002) reflection. The color key shows the $\phi - 2\theta$ angular distribution in the scanned range around the local intensity maxima for the respective axes (c). These maps show the projections from a volume illuminated by box-shaped beam having a size of $ 400 \un{\mu m} \times 400 \un{\mu m}$, rather than a line focused beam (See Supp. Mat. for more details). The measured strain is rather homogeneous having values below $10^{-4}$, in spite of the clear boundaries that are visible in the $\phi$ COM map. }
    \label{fig:strain}
\end{figure}

The DFXM map reveals the self-organization of dislocations upon high-temperature annealing into well-defined planar boundaries, which are only identifiable because of the large field of view and high angular resolution. Due to the high dislocation mobility induced by temperature and the long free paths of motion owing to the low dislocation density, the patterns formed must be the stable preferred configurations. As demonstrated in the analysis above, the observed boundaries are in agreement with low-energy dislocation structures. We confirm the boundary stability by two other DFXM scanning modalities (computed COM maps for lattice tilt and axial strain) over a larger field of view (using a 2$\times$ optical objective at the detector) to compare the strain and orientational populations surrounding the boundaries. 
The misorientation (Fig.~\ref{fig:strain}a) corroborates the boundary positions and the axial strain maps over the same region (Fig.~\ref{fig:strain}b) do not show apparent distortions, indicating them not being associated with long-range stresses along the probed crystallographic direction (correlation shown in Fig.~\ref{fig:strain}c). 

In the present case, each boundary appears to consist of one or two parallel sets of dominant dislocations with only small densities of dislocations of other Burgers vectors. This is in contrast to typical findings after plastic deformation where boundaries typically consist of crossing dislocations \cite{McCabe2004,Wei2011,Winther2015}. The preference for parallel dislocation lines after high-temperature annealing may originate from a higher mobility of such boundaries compared to dislocations in a grid. 

A second important difference from deformation-induced planar dislocation boundaries is that we do not see a large number of parallel boundaries with a spacing of the order of micrometers. Here, we observe only a few boundaries in a volume spanning hundreds of micrometers, with arrangements that are not parallel. This is evidence that assembly into a single boundary is energetically favorable compared to the accumulation of dislocations into two similar boundaries with lower dislocation densities and misorientation, which is also the theoretical expectation \cite{Read_Shockley1950}. This phenomenon is again enabled by the high dislocation mobility and the long free paths of motion. In the context of plastic deformation, this also implies that there is an underlying systematic microscopic deformation process which generates and stabilises the evolution of parallel boundaries. The differences between the present well-annealed low dislocation density observations and those in deformed materials form important input to the ongoing research in the field of modeling crystal deformation at the mesoscale using dislocation dynamics \cite{Devincre2011, Sills2016, Starkey2020}.

Comparing the presently observed structures to the boundaries formed in well-annealed polycrystals, it is interesting that the boundaries do not adjoin into a classical triple-junction with 120\degree{} between the boundaries. The 120\degree{} angle is the equilibrium value for boundaries with  equal energies and isotropic energies according to the Herring relation \cite{Herring1951,King2010}. 
As the dislocation densities in the presently observed adjoining dislocation boundaries in the triple junction B1, B1$'$, and B1$''$ are roughly the same, the preference for boundaries in which the dislocations mutually screen each other's stress fields must control the energy and thus the triple junction angles. This is also in agreement with the findings above that the planar boundaries fulfil the Frank equation. The observed change from a planar boundary to a curved one near the triple-junction, however, also indicates an energetically favorable process induced by the junction itself. Interestingly, the boundaries at the triple junction separate along a direction nearly aligned with the dislocation line vectors, as shown in Fig.~\ref{fig:2DProjection}c.


    
Beneath the scale of the individual dislocations observed in this work, it is important to consider that for this type of low-energy dislocation structure persist after 10 hours of annealing at $0.9 T_m$, there must be additional high-energy immobilizing defects inside the sample that immobilize some of the dislocations, causing the energy for annihilation to be higher than the 65 meV ($k_BT$ at 590 $^{\circ}$C). These sub-resolution high-energy structures likely indicate impurities or interstitials that pin individual dislocations, causing the remaining ones to organize into the most stable packing configuration to minimize the strain energy (via stress screening). As such, this work demonstrates the hierarchical structures that are key to understanding the overall structure of the system, even including defects below our resolution and contrast mechanism. One example of this is the the long dislocation that spans from the midsection down to an almost completely formed loop at $(40,40,40)$ (Fig.~\ref{fig:3Ddislocations}, which is far beyond the initial curved boundary).

\section{Conclusions}

The present study demonstrates the dislocation structures that persist in a bulk single crystal of aluminum after high-temperature annealing. Using dimensional-reduction algorithms on a 4D DFXM dataset (2D scans), we measure the hierarchies of dislocations that span boundaries over hundreds of micrometers within a volume of $100\times300\times300$ $\mu $m$^3$, resolving the boundary planes and vectors of the component dislocations. Our results indicate the self-organization of dislocations into well-defined planar boundaries that separate sub-grains, whose mean dislocation spacings match the misorientations between subdomains that were measured independently with mosaicity maps, despite the heterogeneity. 

Based on the high-dislocation mobility afforded by the high annealing temperature that gave rise to this structure, we conclude that the few but long and coherent boundaries that persist indicate that forming single boundaries is energetically favored as compared to forming two boundaries with lower dislocation densities. It was further concluded that the dislocation configuration in the boundaries is in agreement with theoretical low-energy dislocation structures free of long-range stresses and that triple junction angles are strongly influenced by the resulting anisotropy of the boundary energy.
Our results provide unprecedented information about dislocation patterning in bulk volumes, opening up new avenues not only for potential future experiments to study crystal plasticity but also for new input parameters for modelling. Furthermore, by mapping multiple Bragg reflections, full strain and orientation tensors of individual dislocations in the boundaries can be determined. Our work in this direction is in progress.


\section*{Acknowledgement}
We thank the ESRF for provision of beamtime at ID06-HXM. GW and HFP acknowledge support  from  ERC Advanced Grant nr 885022 and  from  the Danish ESS lighthouse on hard materials in 3D, SOLID. LDM's initial contributions on the experiment were performed under the auspices of the U.S. Department of Energy by Lawrence Livermore National Laboratory under Contract DE-AC52-07NA27344, and the Lawrence Fellowship.

\nocite{*}
\bibliography{extensive}

\begin{thebibliography}{10}
\expandafter\ifx\csname url\endcsname\relax
  \def\url#1{\texttt{#1}}\fi
\expandafter\ifx\csname urlprefix\endcsname\relax\def\urlprefix{URL }\fi
\expandafter\ifx\csname href\endcsname\relax
  \def\href#1#2{#2} \def\path#1{#1}\fi

\bibitem{Orowan1934}
E.~Orowan, Die mechanischen {F}estigkeitseigenschaften und die {R}ealstruktur
  der {K}ristalle, Zeitschrift F{\"u}r Krist. - Cryst. Mater. 89 (1934)
  327--343.

\bibitem{Taylor1934}
G.~I. Taylor, The theory of plasticity of crystals, Zeitschrift F{\"u}r Krist.
  - Cryst. Mater. 89 (1934) 375--385.

\bibitem{Polaniy1934}
M.~Polanyi, About a grid disturbance, which could make a crystal plastic, Z.
  Phys. 89 (1934) 660.

\bibitem{Anderson2017}
P.~M. Anderson, J.~P. Hirth, J.~Lothe, Theory of Dislocations., Cambridge
  University Press, 2017.

\bibitem{Godfrey2000}
A.~Godfrey, D.~A. Hughes, Scaling of the spacing of deformation induced
  dislocation boundaries, Acta Materialia 48 (2000) 1897--105.

\bibitem{Hughes1998}
D.~A. Hughes, D.~C. Chrzan, Q.~Liu, N.~Hansen, Scaling of misorientation angle
  distributions, Phys. Rev. Lett. 81 (1998) 4664--4667.

\bibitem{Winther2007}
G.~Winther, X.~Huang, Dislocation structures. {P}art {II}. {S}lip system
  dependence, Philos. Mag 87 (2007) 5215--5235.

\bibitem{Le2012}
G.~M. Le, A.~Godfrey, C.~S. Hong, X.~Huang, G.~Winther, Dislocation structures.
  {P}art {II}. {S}lip system dependence, Scripta Materialia 66 (2012) 359--362.

\bibitem{McCabe2004}
R.~McCabe, A.~Misra, T.~Mitchell, Experimentally determined content of a
  geometrically necessary dislocation boundary in copper, Acta Materialia 52
  (2004) 705--714.

\bibitem{Wei2011}
Y.~Wei, A.~Godfrey, W.~Liu, Q.~Liu, X.~Huang, N.~Hansen, G.~Winther, Low-energy
  dislocation structure (leds) character of dislocation boundaries aligned with
  slip planes in rolled aluminium, Scr. Mater. 65 (2011) 355--358.

\bibitem{Winther2015}
G.~Winther, C.~Hong, X.~Huang, Low-energy dislocation structure (leds)
  character of dislocation boundaries aligned with slip planes in rolled
  aluminium, Philos. Mag. 95 (2015) 1471--1489.

\bibitem{hata2020electron}
S.~Hata, T.~Honda, H.~Saito, M.~Mitsuhara, T.~Petersen, M.~Murayama, Electron
  tomography: An imaging method for materials deformation dynamics, Current
  Opinion in Solid State and Materials Science 24~(4) (2020) 100850.

\bibitem{feng2020tem}
Z.~Feng, R.~Fu, C.~Lin, G.~Wu, T.~Huang, L.~Zhang, X.~Huang, Tem-based
  dislocation tomography: Challenges and opportunities, Current Opinion in
  Solid State and Materials Science 24~(3) (2020) 100833.

\bibitem{Simons2015}
H.~Simons, A.~King, W.~Ludwig, C.~Detlefs, W.~Pantleon, S.~Schmidt,
  F.~St{\"o}hr, I.~Snigireva, A.~Snigirev, H.~F. Poulsen, Dark field x-ray
  microscopy for multiscale structural characterization, Nat. Commun. 6 (2015)
  6098.
\newblock \href {https://doi.org/10.1038/ncomms7098}
  {\path{doi:10.1038/ncomms7098}}.

\bibitem{Poulsen2020}
H.~F. Poulsen, Multi scale hard x-ray microscopy, Current Opinion in Solid
  State and Materials Science 24 (2020) 100820.
\newblock \href {https://doi.org/10.1016/j.cossms.2020.100820}
  {\path{doi:10.1016/j.cossms.2020.100820}}.

\bibitem{Tanner1976}
B.~K. Tanner, Diffraction Topography., Pergamon Press, 1976.

\bibitem{Ludwig2001}
W.~Ludwig, P.~Cloetens, J.~H{\"a}rtwig, J.~Baruchel, B.~Hamelin, P.~Bastie,
  Three-dimensional imaging of crystal defects by `topo-tomography', J. Appl.
  Crystallogr. 34 (2001) 602--607.
\newblock \href {https://doi.org/10.1107/S002188980101086X}
  {\path{doi:10.1107/S002188980101086X}}.

\bibitem{Hanschke2012}
D.~H{\"a}nschke, L.~Helfen, V.~Altapova, A.~Danilewsky, T.~Baumbach,
  Three-dimensional imaging of dislocations by x-ray diffraction laminography,
  Appl. Phys. Lett. 101 (2012) 244103.
\newblock \href {https://doi.org/10.1063/1.4769988}
  {\path{doi:10.1063/1.4769988}}.

\bibitem{Ahl2017}
S.~R. Ahl, H.~Simons, Y.~B. Zhang, C.~Detlefs, F.~St\"{o}hr, A.~C. Jakobsen,
  D.~Juul~Jensen, H.~F. Poulsen, Ultra-low-angle boundary networks within
  recrystallizing grains, Scripta Mater. 139 (2017) 87--91.
\newblock \href {https://doi.org/10.1016/j.scriptamat.2017.06.016}
  {\path{doi:10.1016/j.scriptamat.2017.06.016}}.

\bibitem{Mavrikakis2019}
N.~Mavrikakis, C.~Detlefs, P.~K. Cook, M.~Kutsal, A.~P.~C. Campos, M.~Gauvin,
  P.~R. Calvillo, W.~Saikaly, R.~Hubert, H.~F. Poulsen, A.~Vaugeois,
  H.~Zapolsky, D.~Mangelinck, M.~Dumont, C.~Yildirim, A multi-scale study of
  the interaction of {S}n solutes with dislocations during static recovery in
  $\alpha$-{F}e, Acta Mater. 174 (2019) 92--104.
\newblock \href {https://doi.org/10.1016/j.actamat.2019.05.021}
  {\path{doi:10.1016/j.actamat.2019.05.021}}.

\bibitem{yildirim20224d}
C.~Yildirim, N.~Mavrikakis, P.~Cook, R.~Rodriguez-Lamas, M.~Kutsal, H.~Poulsen,
  C.~Detlefs, 4d microstructural evolution in a heavily deformed ferritic
  alloy: A new perspective in recrystallisation studies, Scripta Materialia 214
  (2022) 114689.

\bibitem{Simons2018}
H.~Simons, A.~Haugen, A.~Jakobsen, S.~Schmidt, F.~St{\"o}hr, M.~Majkut,
  C.~Detlefs, J.~Daniels, D.~Damjanovic, H.~Poulsen, Long-range symmetry
  breaking in embedded ferroelectrics, Nat. Mater. 17 (2018) 814--819.
\newblock \href {https://doi.org/10.1038/s41563-018-0116-3}
  {\path{doi:10.1038/s41563-018-0116-3}}.

\bibitem{Bucsek2019}
A.~Bucsek, H.~Seiner, H.~Simons, P.~Cook, C.~Yildirim, Y.~Chumlyakov,
  C.~Detlefs, A.~P. Stebner, Sub-surface measurements of the austenite
  microstructure in response to martensitic phase transformation, Acta Mater.
  179 (2019) 273--286.
\newblock \href {https://doi.org/10.1016/j.actamat.2019.08.036}
  {\path{doi:10.1016/j.actamat.2019.08.036}}.

\bibitem{Jakobsen2019}
A.~C. Jakobsen, H.~Simons, W.~Ludwig, C.~Yildirim, H.~Leemreize, L.~Porz,
  C.~Detlefs, H.~F. Poulsen, Mapping of individual dislocations with dark-field
  x-ray microscopy, J. Appl. Cryst. 52 (2019) 122.
\newblock \href {https://doi.org/10.1107/S1600576718017302}
  {\path{doi:10.1107/S1600576718017302}}.

\bibitem{Simons2019}
H.~Simons, A.~C. Jakobsen, S.~R. Ahl, H.~F. Poulsen, W.~Pantleon, Y.-H. Chu,
  C.~Detlefs, N.~Valanoor, Nondestructive mapping of long-range dislocation
  strain fields in an epitaxial complex metal oxide, Nano Lett. 19 (2019)
  1445--1450.
\newblock \href {https://doi.org/10.1021/acs.nanolett.8b03839}
  {\path{doi:10.1021/acs.nanolett.8b03839}}.

\bibitem{cockayne1969}
D.~Cockayne, I.~Ray, M.~Whelan, Investigations of dislocation strain fields
  using weak beams, Philosophical Magazine 20~(168) (1969) 1265--1270.

\bibitem{humphreys_2004}
F.~J. Humphreys, M.~Hatherly, Recrystallization {And} {Related} {Annealing}
  {Phenomena}, 2nd Edition, Elsevier, 2004.
\newblock \href {https://doi.org/10.1016/B978-0-08-044164-1.X5000-2}
  {\path{doi:10.1016/B978-0-08-044164-1.X5000-2}}.

\bibitem{Kutsal2019}
M.~Kutsal, P.~Bernard, G.~Berruyer, P.~K. Cook, R.~Hino, A.~C. Jakobsen,
  W.~Ludwig, J.~Ormstrup, T.~Roth, H.~Simons, K.~Smets, J.~X. Sierra, J.~Wade,
  P.~Wattecamps, C.~Yildirim, H.~F. Poulsen, C.~Detlefs, The esrf dark-field
  x-ray microscope at id06, IOP Conf. Series: Materials Science and Engineering
  580 (2019) 012007.
\newblock \href {https://doi.org/10.1088/1757-899X/580/1/012007}
  {\path{doi:10.1088/1757-899X/580/1/012007}}.

\bibitem{Poulsen2021}
H.~F. Poulsen, L.~E. Dresselhaus-Marais, M.~A. Carlsen, C.~Detlefs, G.~Winther,
  {Geometrical Optics Formalism to Model Contrast in Dark-Field X-ray
  Microscopy}, J. Appl. Crystal. 54 (2021) 1555--1571.

\bibitem{darfix}
J.~G. Ferrer, et~al., darfix: Data analysis for dark-field x-ray microscopy,
  arXiv:2205.05494 (2022).

\bibitem{SurfTool}
A.~Adam,
  \href{https://www.mathworks.com/matlabcentral/fileexchange/28497-plot-a-3d-array-using-patch}{Plot
  a 3d array using patch} (2021).
\newline\urlprefix\url{https://www.mathworks.com/matlabcentral/fileexchange/28497-plot-a-3d-array-using-patch}

\bibitem{Hoppe1992}
H.~Hoppe, T.~DeRose, T.~Duchamp, J.~McDonald, W.~Stuetzle, Surface
  reconstruction from unorganized points, in: Proceedings of the 19th annual
  conference on computer graphics and interactive techniques, 1992, pp. 71--78.

\bibitem{torr2000mlesac}
P.~H. Torr, A.~Zisserman, Mlesac: A new robust estimator with application to
  estimating image geometry, Computer vision and image understanding 78~(1)
  (2000) 138--156.

\bibitem{Hull2011}
D.~Hull, D.~Bacon, Introduction to Dislocations, 5th Edition, Elsevier Ltd.,
  2011.

\bibitem{Brennan}
M.~Brennan, et~al., Analytical methods for superresolution dislocation identi
  cation in dark-field x-ray microscopy, arXiv:2203.05671 (2022).

\bibitem{Frank1950}
F.~Frank, The resultant content of dislocations in an arbitrary
  intercrystalline boundary, Proc. of Symposium on plastic deformation of
  crystalline solids, Mellon Institute of Industrial Research, Pittsburgh
  (1950) 150--154.

\bibitem{zhang2010high}
J.-S. Zhang, High temperature deformation and fracture of materials, Elsevier,
  2010.

\bibitem{Read_Shockley1950}
W.~Read, W.~Shockley, Dislocation models of crystal grain boundaries, Phys.
  Rev. 78 (1950) 275--289.

\bibitem{Devincre2011}
B.~Devincre, R.~Madec, G.~Monnet, S.~Queyreau, R.~Gatti, L.~Kubin, Modeling
  crystal plasticity with dislocation dynamics simulations: the
  ``micro{M}egas'' code, Mech. Nano-Objects 1 (2011) 81--99.

\bibitem{Sills2016}
R.~Sills, W.~Kuykendall, A.~Aghaei, W.~Cai, Fundamentals of dislocation
  dynamics simulations, Multiscale Materials Modeling for Nanomechanics.
  Springer Series in Materials Science 245 (2016) 53--87.

\bibitem{Starkey2020}
K.~Starkey, G.~Winther, A.~El-Azab, Theoretical development of continuum
  dislocation dynamics for finite-deformation crystal plasticity at the
  mesoscale, Journal of the Mechanics and Physics of Solids 139 (2020) 103926.

\bibitem{Herring1951}
C.~Herring, Surface tension as a movitation for sintering, In The physics of
  powder metallurgy (ed. W.W. Kingston), McGraw-Hill, NY (1951) 143--179.

\bibitem{King2010}
A.~King, Triple lines in materials science and engineering, Scr. Mater. 62
  (2010) 889--893.

\bibitem{Porz2021}
L.~Porz, A.~J. Klomp, X.~N. Li, C.~Yildirim, C.~Detlefs, E.~Bruder,
  M.~H{\"o}fling, W.~Rheinheimer, E.~A. Patterson, P.~Gao, K.~Durst,
  A.~Nakamura, K.~Albe, H.~Simons, J.~R{\"o}del, Dislocation-toughened
  ceramics, Mater. Horizons 8(5) (2021).
\newblock \href {https://doi.org/10.1039/D0MH02033H}
  {\path{doi:10.1039/D0MH02033H}}.

\bibitem{ahl2018}
S.~R. Ahl, Elements of a method for multiscale characterization of
  recrystallization in deformed metals, Ph.D. thesis, Department of Physics,
  Technical University of Denmark (2018).

\bibitem{Ulvestad2017}
A.~Ulvestad, M.~J. Welland, W.~Cha, Y.~Liu, J.~W. Kim, R.~Harder, E.~Maxey,
  J.~N. Clark, M.~J. Highland, H.~You, P.~Zapol, S.~O. Hruszkewycz, G.~B.
  Stephenson, Three-dimensional imaging of dislocation dynamics during the
  hydriding phase transformation, Nat. Mater. 16 (2017) 565--571.

\bibitem{Hofmann2013a}
F.~Hofmann, B.~Abbey, W.~Liu, R.~Xu, B.~F. Usher, E.~Balaur, Y.~Liu, X-ray
  micro-beam characterization of lattice rotations and distortions due to an
  individual dislocation, Nat. Commun. 4 (2013) 2774.
\newblock \href {https://doi.org/10.1038/ncomms3774}
  {\path{doi:10.1038/ncomms3774}}.

\bibitem{Dresselhaus2021}
L.~E. Dresselhaus-Marais, G.~Winther, M.~Howard, A.~Gonzalez, C.~Yildirim,
  P.~K. Cook, M.~Kutsal, L.~Zeppeda-Ruiz, A.~Samanta, C.~Detlefs, J.~H. Eggert,
  H.~Simons, H.~F. Poulsen, In-situ visualization of long-range dislocation
  interactions in the bulk, Sci. Advances 7 (2021) eabe8311.
\newblock \href {https://doi.org/10.1126/sciadv.abe8311}
  {\path{doi:10.1126/sciadv.abe8311}}.

\bibitem{Asadchikov2018}
V.~Asadchikov, A.~Buzmakov, F.~Chukhovskii, I.~Dyachkova, D.~Zolotov,
  A.~Danilewsky, T.~Baumbach, S.~Bode, S.~Haaga, D.~Hanschke, M.~Kabukcuoglu,
  M.~Balzer, M.~Caselle, E.~Suvorov, X-ray topo-tomography studies of linear
  dislocations in silicon single crystals, J. Appl. Cryst 51 (2018) 1616--1622.
\newblock \href {https://doi.org/10.1107/S160057671801419X}
  {\path{doi:10.1107/S160057671801419X}}.

\bibitem{Danilewsky2020}
A.~N. Danilewsky, X-ray topography -- more than nice pictures, Crystal Research
  and Technology 55~(9) (2020) 2000012.
\newblock \href {https://doi.org/https://doi.org/10.1002/crat.202000012}
  {\path{doi:https://doi.org/10.1002/crat.202000012}}.

\bibitem{Gastaldi1988}
J.~Gastaldi, C.~Jourdan, G.~Grange, \textit{In situ} synchrotron x-ray
  topography study of the generation of lattice dislocations in aluminium by
  migrating grain boundaries, Philosophical Magazine A 57 (1988) 971--981.

\end{thebibliography}

\end{document}